\documentclass{scrartcl}
\usepackage{amsmath,amssymb,amsthm}
\usepackage{enumerate}
\usepackage{graphicx}
\usepackage{xcolor}
\usepackage{tikz}
\usetikzlibrary{arrows}
\usepackage{mathtools}
\mathtoolsset{showonlyrefs}

\newtheorem{theorem}{Theorem}[section]
\newtheorem{lemma}[theorem]{Lemma}

\newtheorem{corollary}[theorem]{Corollary}

\theoremstyle{definition}
\newtheorem{definition}[theorem]{Definition}
\theoremstyle{remark}
\newtheorem{remark}[theorem]{Remark}
\newtheorem{example}[theorem]{Example}

\newcommand{\sign}{\mathop{\mathrm{sign}}}
\usepackage{scalerel,stackengine}
\stackMath
\newcommand\reallywidehat[1]{%
\savestack{\tmpbox}{\stretchto{%
\scaleto{%
\scalerel*[\widthof{\ensuremath{#1}}]{\kern-.6pt\bigwedge\kern-.6pt}%
{\rule[-\textheight/2]{1ex}{\textheight}}
}{\textheight}%
}{0.5ex}}%
\stackon[1pt]{#1}{\tmpbox}%
}
\DeclareMathOperator{\reldist}{reldist}

\DeclareMathOperator{\dimension}{dim}
\DeclareMathOperator{\sgn}{sgn}

\begin{document}
\title{Perturbation of eigenvalues of the Klein-Gordon operators}
\author{Ivica Naki\'c\thanks{Department of Mathematics, Faculty of Science, University of Zagreb, e-mail: nakic@math.hr} \and Kre\v simir Veseli\'c\thanks{Fernuniversit\"at Hagen, Fakult\"at f\"ur Mathematik und Informatik, Postfach 940, D-58084 Hagen, Germany, e-mail: Kresimir.Veselic@FernUni-Hagen.de}}

\maketitle

\begin{abstract} 
We prove inclusion theorems for both spectra and essential spectra as well as two-sided bounds for isolated eigenvalues 
for Klein-Gordon type Hamiltonian operators.
We first study operators of the form $JG$, where $J$, $G$ are selfadjoint operators on a Hilbert space,
$J = J^* = J^{-1}$ and $G$ is positive definite and then we apply these results to 
obtain bounds of the Klein-Gordon eigenvalues under the change of the electrostatic potential. The developed general theory allows applications to some other
instances, as e.g.\ the Sturm-Liouville problems with indefinite weight.
\end{abstract}
\section{Introduction and preliminaries} 
\label{sec:setting}

The main object of our considerations will be the abstract Klein-Gordon equation 
\begin{equation}\label{KG}
 \left((i\frac{\partial}{\partial t} - V)^2 - U^2\right)\psi = 0 
\end{equation}
where \(U,\ V\) are operators in a Hilbert space \(\mathcal{X}\)
such that \(U\) is selfadjoint and positive definite and
\(V\) is symmetric.\footnote{
Whenever otherwise specified we shall use the terminology and notation
of \cite{Kt-66}.}
The Klein–Gordon equation is the relativistic version of the Schr\"odinger equation, which
is used to describe spinless particles. We refer the reader to \cite{greiner} for the physical background.

To this end, we study the spectral properties and perturbation of the spectra for the operators on a Hilbert space which have the form $JG$, where $J$ is a reflection i.e.\ $J=J^*= J^{-1}$, and $G$ is positive definite selfadjoint operator. We then apply the obtained results to the case of abstract Klein-Gordon operator. The developed abstract theory is interesting in its own right, which we illustrate with some examples.

In the case of the Klein-Gordon operator, typically, one has
\(\mathcal{X} = \mathbf{L}^2(\mathbb{R}^n)\), \( n = 1,2,3\);
\(U^2 = m^2 - \Delta\) and \(V\psi(x) = 
V(x)\psi(x),\ x \in \mathbb{R}^n\) a multiplication operator; this 
describes a spinless quantum relativistic particle with mass \(m\)
in an external electrostatic field described by \(V\)
(all measured in appropriate units). More generally, \(-\Delta\)
can be replaced by \((-i\nabla - A)^2\) with an external magnetic
potential \(A = A(x)\), and also the constant \(m\) can be replaced by a
positive multiplication operator \(m(x)\). Finally, \(U^2\) and \(V\)
may be finite matrices obtained, for instance, as a discretisation
of an infinite dimensional system. In each case, using the representation based on Feshbach-Villars \cite{feshb}, the Klein-Gordon equation gives rise to a Hamiltonian matrix operator.

Under appropriate conditions the Hamiltonian will 
be selfadjoint with respect to a new scalar product which is topologically
equivalent to the original one, but the latter scalar product {\em changes
with the potential} and this will be the main difficulty to cope with.

Our aim is to study the changes of its eigenvalues caused by the changes of the
potential \(V\).
To this end, we linearise \eqref{KG} in a particular manner which enables us to use the obtained abstract perturbation results. Our linearisation is not just a mathematical tool but also has a physical meaning, being a generalisation of the so--called number norm.

The abstract Klein-Gordon equation in an operator-theoretic setting similar as in this paper has already been extensively studied, see \cite{Jonas2000,LangerNajmanTretter2006,LangerNajmanTretter2008,LangerTretter2006,Koppen2011} and references therein. Study of eigenvalues of the Klein-Gordon operator as an operator in an indefinite inner product space has an even longer history, starting with the paper \cite{v0old} and continuing with \cite{lundberg1973,Kako1976,najman1980spectral,Najman1980,Najman1983,v0new,Jonas1993}, a list which is far from being complete.
However, to the best of our knowledge, the perturbation theory for the eigenvalues of the Klein-Gordon operator is not so well studied, especially the relative bounds for the perturbation of the eigenvalues were not previously known.  

The main assumptions on the operators \(U,V\) above are
\begin{enumerate}
	\item[A.1] $U$ is a positive definite\footnote{In this paper positive definiteness means strict positivity: there exists $\delta > 0$ such that $(U \psi, \psi)\ge \delta (\psi,\psi)$ for all $\psi\in \mathcal{D}(U)$.  We obtain positive semi-definiteness if we assume $\delta \ge 0$.} selfadjoint operator on a Hilbert space $\mathcal{X}$,
	\item[A.2] $V$ is a symmetric operator on $\mathcal{X}$ with $\mathcal{D}(V) \supset \mathcal{D}(U)$, and
	\item[A.3] there exists $\mu\in \mathbf{R}$ such that 
\begin{equation}
\label{muVb}
\lVert (\mu I- V)U^{-1} \rVert =b <1.
\end{equation}
\end{enumerate}
Typical estimates we obtain look like
\[
|\delta \lambda| \leq c\|\delta V\|,\quad c \geq 1,
\]
and 
\[
\frac{|\delta \lambda|}{\lvert \lambda \rvert } \leq c\|\delta V\, U^{-1}\|,\quad c \geq 1,
\]
where \(\delta \lambda, \delta V\) are the changes of the eigenvalue \(\lambda\)
and the potential, \(V\), respectively. One example of such an estimate is \eqref{lam_lam'1}.
Unlike in standard selfadjoint situations here the constant cannot in general 
 be pushed
down to one.\footnote{There are indeed important special classes of potentials and perturbations for which
the constant \(c\) will be equal to one. Such estimates are obtained by a detailed study of
the quadratic eigenvalue equation attached to \eqref{KG}; this should be presented in a subsequent paper.} 
The estimates are obtained from minimax formulae, so they hold for the (properly ordered) 
eigenvalues with their multiplicities. 

It should be noted that most
of the results in this paper are novel even in the finite dimensional case.

Section 2 collects mostly known results: the construction of the Hamiltonian
and its spectral theory based on its selfadjointness and positivity in a
canonically given Krein space environment. Section 3 presents inclusion theorem for the spectra as well as
eigenvalue estimates for a class of quasi-Hermitian operators. 
The methods used here are partially inspired by \cite{vs} which dealt with finite matrices. 
In Section 4 we apply the results of Section 3 to the special operator
matrix structure of the Klein-Gordon Hamiltonian. Finally, Section 5 gives a
selection of illustrating examples for both the Klein-Gordon equation as well as the Sturm-Liouville eigenvalue problem.
\section{The Hamiltonian operator} 
\label{sub:operator_construction}
By the formal substitution
\begin{equation}\label{linearisation}
\psi_1 = U^{1/2}\psi,\quad \psi_2 = U^{-1/2}
\left(i\frac{\partial}{\partial t} - V\right)\psi
\end{equation}
\eqref{KG} goes over to
\begin{equation}\label{KG_op}
\frac{\partial}{\partial t}
\begin{bmatrix}
\psi_1  \\
\psi_2 \\
\end{bmatrix}
=
\mathbf{H}
\begin{bmatrix}
\psi_1 \\
\psi_2 \\
\end{bmatrix},\quad
\mathbf{H} =
\begin{bmatrix}
U^{1/2}VU^{-1/2} &          U             \\
 U                      &U^{-1/2}VU^{1/2} \\
\end{bmatrix}
\end{equation}
Here \(\mathbf{H}\) is called {\em the abstract Klein-Gordon operator} or 
{\em abstract Klein-Gordon Hamiltonian}. This conversion of a second order
equation into a system of first order (oft called 'the linearisation') 
is here not merely the means
of studying this equation. In fact, quantum mechanical interpretation
needs a first order equation in Hamiltonian form as was established
already in \cite{feshb}. Our linearisation 
exhibits a particular 'weighting' implicitly present in \cite{v0old};
in this way the Hilbert space norm used here is a
direct generalisation of the
so-called 'number norm' introduced in \cite{Seiler}.\footnote{The term 
'number norm' stems from the fact that upon second quantisation the
number norm gives rise to the number of particles operator.}

Introducing the fundamental symmetry
\begin{equation}\label{JJ}
\mathbf{J} =
\begin{bmatrix}
 0   & I    \\
I    & 0 \\
\end{bmatrix}
\end{equation}
we see that \(\mathbf{J}\mathbf{H}\) becomes symmetric: 
\begin{equation}\label{GG}
\mathbf{G} = \mathbf{J}\mathbf{H} =
\begin{bmatrix}
 U               &U^{-1/2}VU^{1/2}\\
U^{1/2}VU^{-1/2} &          U \\
\end{bmatrix}.
\end{equation}
In other words,
\(\mathbf{H}\) is symmetric with respect to the indefinite form
generated by \(\mathbf{J}\). The same is true of the 
{\em free Hamiltonian}
\begin{equation}\label{HH0}
\mathbf{H}_0 =
\begin{bmatrix}
 0 &   U \\
 U &   0 \\
\end{bmatrix}
\end{equation}
which is also selfadjoint in the standard scalar product
and
\[
\mathbf{J}_0 = \sign\mathbf{H}_0.
\]
All this enables us to construct a bona fide Hamiltonian by using
forms.

The operator \(\mathbf{G}\) is naturally given by the form sum\footnote{We
will freely use sums of operators and forms by understanding them
to be taken in the form sense.}
\[
\mathfrak{g} = \mathfrak{g}_0 + \mathfrak{v}
\]
where
\begin{equation*}
\mathfrak{g}_0\left(\left(\begin{smallmatrix}
  \psi_1 \\ \psi_2
\end{smallmatrix}\right),\left(\begin{smallmatrix}
  \varphi_1 \\ \varphi_2
\end{smallmatrix}\right)\right) = 
(U^{1/2} \psi_1,U^{1/2}\varphi_1) + (U^{1/2} \psi_2, U^{1/2}\varphi_2)
\end{equation*} 
belongs to the selfadjoint positive definite operator
\begin{equation}\label{GG0}
\mathbf{U} =
\begin{bmatrix}
U   & 0  \\
0   & U  \\
\end{bmatrix}
= \mathbf{J}\mathbf{H_0} = |\mathbf{H}_0|.
\end{equation}
The form
\begin{equation}\label{vv}
\mathfrak{v}\left(\left(\begin{smallmatrix}
  \psi_1 \\ \psi_2
\end{smallmatrix}\right),\left(\begin{smallmatrix}
  \varphi_1 \\ \varphi_2
\end{smallmatrix}\right)\right) = (U^{1/2}\psi_2,VU^{-1/2}\varphi_1)  + 
 (VU^{-1/2} \psi_1, U^{1/2}\varphi_2)
\end{equation}
is symmetric and defined on \(\mathcal{D}(\mathbf{U}^{1/2})\)
The condition A.3 immediately implies
\begin{equation}\label{vv<bU}
|\mathfrak{v}(\psi,\psi)- \mu(\mathbf{J}\psi,\psi)| \leq 
b(\mathbf{U}^{1/2}\psi,\mathbf{U}^{1/2}\psi),\quad b < 1,
\end{equation}
so, by the standard theory (see \cite[Ch.~VI]{Kt-66}) \(\mathfrak{g} - \mu J\) 
generates 
a positive definite operator \(\mathbf{G} - \mu\mathbf{J}\) with 
\[
\mathcal{D}((\mathbf{G} - \mu\mathbf{J})^{1/2} 
= \mathcal{D}(\mathbf{U}^{1/2}). 
\]
This operator can be written as
\begin{equation}\label{G_factor}
\mathbf{G} - \mu\mathbf{J} = \mathbf{U}^{1/2}\mathbf{A}\mathbf{U}^{1/2}
\end{equation}
where
\begin{equation}\label{bfA}
\mathbf{A} =   
\begin{bmatrix}
I & A^{\ast} \\
A & I
\end{bmatrix}
\end{equation}
and  $A = (V - \mu) U^{-1}$, $\|A\| \leq b$. Note that
\(\mathbf{A}\) is bounded positive definite with
\[
\|\mathbf{A}^{-1}\| \leq \frac{1}{1 - b}.
\]

By setting \(\mathbf{H} =\mathbf{J} \mathbf{G}\) we have given 
rigorous meaning to the operators \(\mathbf{H},\mathbf{G}\) introduced
formally in the previous section.
The operator $\mathbf{H}$ is $\mathbf{J}$--selfadjoint, 
that is, it is selfadjoint in the Krein space with the fundamental 
symmetry given by $\mathbf{J}$, in other words, 
\(\mathbf{H}^* = \mathbf{J}\mathbf{H}\mathbf{J}\). Another way to express the same property is to say that $\mathbf{H}$ is pseudo-Hermitian (c.f. \cite{albeverio}) with the intertwining operator $\mathbf{J}$. 

We want to analyse spectral properties of operators which have the form $H = J G$, where $J$ is a reflection ($J = J^* = J^{-1}$) and $G$ is a positive definite operator. Note that $H$ will have such a form if and only if the unitary part of its polar decomposition is a reflection. Similarity of operators will play a crucial role in the paper. We say that two operators $A$ and $B$ are \emph{similar} if there exists a bounded and boundedly invertible operator $T$ such that $B = T^{-1}A T$.

The following theorem will play an important role in this paper.
\begin{theorem}\label{spectrality}
Let $H_0$ be closed operator such that $0\in \rho(H_0)$ and such that the unitary part $J$ of its polar decomposition $H_0 = J G_0 $ is a reflection. 
Let \(v\), $\mathcal{D}(v)\supseteq \mathcal{D}(G_0^{1/2})$ be a symmetric
form satisfying the inequality
\begin{equation}\label{G0vb}
|v(\psi,\psi) - \mu (J \psi, \psi) | 
\leq b (G_0^{1/2}\psi,G_0^{1/2}\psi ),\quad \text{ for all }\psi \in \mathcal{D}(G_0^{1/2}),
\end{equation}
with $b < 1$ for some real $\mu$, and let \( G - \mu J\) be the
selfadjoint positive definite  operator generated by the form
sum \(G_0  + v - \mu J\). Then $H = J G$ is similar to a selfadjoint operator if and only if $H_0$ is similar to a selfadjoint operator.
\end{theorem}
\begin{proof}
If we equip the underlying Hilbert space $\mathcal{X}$ with the indefinite scalar product given by $J$, $H_0$ and $H$ become positive selfadjoint operators in the corresponding Krein space. The proof now follows from \cite[Theorem 2.5]{Curgus85} and 
\cite[Proposition 2.1]{CurgusNajman95} (see also \cite{LangerNajmanTretter2008}).
 \end{proof}
\begin{remark}
The main direction of the theorem can also be proved using \cite{v0}. There the form $v$ is represented by an operator and $\mu = 0$ but the proofs are 
unchanged with $v$ as a form and $\mu$ is just a spectral shift. The main direction of the result
has also been proved in 
is also proved in \cite{vs3} 
in the special case of the Klein-Gordon operator by a different method.
\end{remark}
\begin{remark}
	Note that the polar decomposition of the operator $H$ from the preceding theorem is given by $H = J G$. If $H_0$ is selfadjoint than the preceding implies that $H$ is similar to a selfadjoint operator, which will be the case for our Klein-Gordon operator. 
\end{remark}
Since operators similar to selfadjoint operators are called quasi-Hermitian in the physics literature (c.f.\ \cite{scholtz1992quasi}), we will 
adopt the following definition.
\begin{definition}
An operator will be called $j$-\textit{quasi-Hermitian}, if
\begin{itemize}
\item it is similar to a selfadjoint operator,
\item it has zero in its resolvent set,
\item the unitary part of its polar decomposition is a reflection.
\end{itemize}	
\end{definition}
A most notorious example satisfying the conditions of the preceding theorem
is the standard Coulomb Hamiltonian in \(\mathcal{X} = L^2(\mathbb{R}^3)\), 
$U^2 = - \Delta + m^2$ and $V = V(x) = \zeta/|x|$ where by the well known 
Hardy bound (\cite{Kt-66}, Ch.~V.5.4) we have the sharp estimate 
\begin{equation}\label{coulomb}
b = \zeta/2. 
\end{equation}
%
It is easy to see that the spectrum \(\sigma(H)\) to the right/left  of \(\mu\) is characterised by the property $ (J\psi,\psi)> 0$/$(J\psi,\psi) < 0$  for all non-vanishing \(\psi\) from the respective spectral subspace and that the whole Hilbert space is a $J$-orthogonal sum of the two mentioned subspaces.
This property, in fact, shows that the choice of the Hilbert space metric, implicitly made in the linearisation \eqref{linearisation} is canonical. 

The following result has been established in \cite{LangerNajmanTretter2006}. For the sake of the reader, we present a short proof. 
\begin{corollary}
        \label{cor:gap} Assume that we are under the conditions of the preceding 
theorem and that additionally $H_0$ is selfadjoint. Then we have
\begin{equation}\label{gap}
\sigma(H) \cap (\mu -\alpha,\mu + \alpha) = 
\emptyset \mbox{ where } \alpha = (1-b) \inf|\sigma(H_0)|
\end{equation}
\end{corollary}
\begin{proof} By \eqref{G0vb} we have the form inequality
\begin{equation}\label{GleqG0}
G - \mu J \geq (1 - b) G_0
\end{equation}
which implies 
\[
\frac{1}{\inf|\sigma(H) - \mu|} \leq \|(H - \mu)^{-1}\| \le \|(G - \mu J)^{-1}\| 
\leq \frac{\|G_0^{-1}\|}{1 - b} = \frac{\|H_0^{-1}\|}{1 - b} = \frac{1}{\alpha},
\]
which 
then implies \eqref{gap}. 
 \end{proof}

Now, both Theorem \ref{spectrality} and Corollary 
\ref{cor:gap} immediately apply to
our Klein-Gordon operators, just identify 
\(\mathbf{H}_0, \mathbf{H},\mathbf{U},\mathbf{G},\mathbf{J}\)
with \(H_0, H,G_0,G,J\), respectively. In this case the condition A.3 is equivalent
to \eqref{G0vb} and
\[
\inf|\sigma(H_0)| = \inf\sigma(U).
\]
Since $\mathbf{H}_0$ is selfadjoint, it follows that $\mathbf{H}$ is $j$-quasi-Hermitian. 
\begin{remark}\label{shift}
Replacing \(\mu\) by zero in Theorem 
\ref{spectrality} - Corollary \ref{cor:gap} above just means
a spectral shift in \(H\).
More precisely, by explicitly denoting the dependence
on \(v\) as \(H = H_v\) we have
\[
 H_{v - \mu J} = H_v - \mu I,\footnote{Note that \(v - \mu J\)  
corresponds to \(V - \mu I\) in \eqref{muVb}.}
\]
so we will in the following always work with 
\(\mu = 0\) to facilitate the notations. In applications,
however, we will just take a shift making the bound \(b\)
as small as possible.
\end{remark}
For quantitative spectral analysis it is important to establish
the similarity with a \emph{concrete} selfadjoint operator. Now, if
\(G\) (and then also \(H\)) from Theorem \ref{spectrality} is bounded the sought operator is 
simply \(G^{1/2}JG^{1/2}\); indeed,
\[
H = G^{-1/2}G^{1/2}JG^{1/2}G^{1/2}.
\]
This ceases to hold in the unbounded case, the mere 
positive definiteness of \(G\) does not suffice
for the purpose. But in our particular case this is again true; we have the following result.
\begin{lemma}\label{lem:similarity}
Let $H$ be $j$-quasi-Hermitian with its polar decomposition \(H = JG\). Then the operators \(H\) and 
\(G^{1/2}JG^{1/2}\) are similar. 
\end{lemma}
\begin{proof}
Since  \(H\) is similar with some selfadjoint operator \(T\), so are
the inverses which are bounded:
\[
T^{-1} = S^{-1}H^{-1}S = S^{-1}G^{-1}JS
\]
and we arrive at the so called quasi-affinity relation 
\[
T^{-1}S^{-1}G^{-1/2}
= S^{-1}G^{-1/2}G^{-1/2}JG^{-1/2}.
\]
between two bounded selfadjoint operators \(T^{-1}, G^{-1/2}JG^{-1/2}\).
Since \(S^{-1}G^{-1/2} \) as well as its adjoint are injective, 
its polar decomposition \(S^{-1}G^{-1/2} = UP\) 
will have a unitary \(U\) and \(P = P^{\ast} \) will be positive semi-definite
and injective. Hence $U^* T^{-1} U = P G^{-1/2}JG^{-1/2} P^{-1}$. Since $U^* T^{-1} U$ is selfadjoint, it follows \\ $P G^{-1/2}JG^{-1/2} P^{-1} = P^{-1} G^{-1/2}JG^{-1/2} P$, hence $P^2 G^{-1/2}JG^{-1/2} = G^{-1/2}JG^{-1/2} P^2$. This implies that $P$ and $G^{-1/2}JG^{-1/2}$ commute, hence $U^* T^{-1} U = G^{-1/2}JG^{-1/2}$, from which the claim follows immediately. 
 \end{proof}

Thus, the spectral properties of \(H\) are identical to 
those of \(G^{1/2}JG^{1/2}\), which is the crucial fact exploited in the next section. 
\section{Perturbations of spectra, general theory} 
\label{sec:perturbations_of_spectra}

This is the central part of this paper. We first prove inclusion
theorems for both spectra and essential spectra and then two-sided
bounds for isolated eigenvalues as they are usually 
obtained by minimax techniques.

The class of $j$-quasi-Hermitian operators is stable under perturbations.
\begin{theorem}
        \label{stability}
Let $H$ be $j$-quasi-Hermitian with its polar decomposition \(H = JG\) and let \(H' = JG'\) where
\(G'\) is given by the form sum
\begin{equation}\label{deltag}
g' = g + \delta g \mbox{ with }
|\delta g| \leq \kappa g,\quad \kappa < 1,
\end{equation}
and where $\delta g$ is a symmetric form satisfying $\mathcal{D}(\delta g) \supseteq \mathcal{D}(g)$.
Then \(H'\) is again $j$-quasi-Hermitian.
\end{theorem}
\begin{proof}
	Apply Theorem \ref{spectrality} with $H_0 = H$ and \eqref{deltag} playing the role of \eqref{G0vb}.
 \end{proof}
\begin{theorem}\label{theorem:inclu}
Assume that we are under the assumption of the preceding theorem.
Let $(\lambda^-,\lambda^+)$ be a spectral gap for $H$, i.e.\ 
$(\lambda^-,\lambda^+) \subseteq \rho(H)$.
Then we have
\begin{enumerate}[(i)]
  \item $((1+\kappa)\lambda^-,(1 - \kappa)\lambda^+)\subset \rho(H')$ if $\lambda^- >0$,
  \item $((1-\kappa)\lambda^-,(1 - \kappa)\lambda^+)\subset \rho(H')$ if $\lambda^- <0$  and $\lambda^+ >0$, 
  \item $((1-\kappa)\lambda^-,(1 + \kappa)\lambda^+)\subset \rho(H')$ if $\lambda^+ <0$.
\end{enumerate}
Let $(\lambda^-,\lambda^+)$ be an essential spectral gap for $H$, i.e.\ $(\lambda^-,\lambda^+)\cap \sigma_{\mathrm{ess}}(H)=\emptyset$. Then we have
\begin{enumerate}[(i)]
\setcounter{enumi}{3}
  \item $((1+\kappa)\lambda^-,(1 - \kappa)\lambda^+)\cap \sigma_{\mathrm{ess}}(H')=\emptyset$ if $\lambda^- >0$,
  \item $((1-\kappa)\lambda^-,(1 - \kappa)\lambda^+)\cap \sigma_{\mathrm{ess}}(H')=\emptyset$ if $\lambda^- <0$  and $\lambda^+ >0$, 
  \item $((1-\kappa)\lambda^-,(1 + \kappa)\lambda^+)\cap \sigma_{\mathrm{ess}}(H')=\emptyset$ if $\lambda^+ <0$.
\end{enumerate}
\end{theorem}
\begin{proof}
First we note that $g'$ can be written as
\[
g'(\psi,\phi) = ((I + \Delta)G^{1/2}\psi,G^{1/2}\psi) 
\]
where \(\Delta\) is a bounded operator defined by
\begin{equation}\label{Delta}
(\Delta\psi,\phi) =  \delta g(G^{-1/2}\psi,G^{-1/2}\phi)
\end{equation}
and satisfies $\|\Delta\| \leq \kappa$.
From Theorem \ref{stability} it follows that \(G'\) 
defines a positive definite operator $G'$ and $H'$ 
is again $j$-quasi-Hermitian, and we obviously have 
\begin{equation}\label{G'}
G' = G^{1/2}(I + \Delta) G^{1/2}.
\end{equation}
For any \(\lambda\in \mathbb{C}\) the operator \(G' - \lambda J\)
is generated by the closed sectorial form
\[
g'(\psi,\phi) -\lambda (J\psi,\phi) = (ZG^{1/2}\psi,G^{1/2}\phi)
\]
with
\[
Z = I - \lambda G^{-1/2}JG^{-1/2} + \Delta \in \mathcal{B}(\mathcal{X}).
\]
Let us define the {\em relative distance} of the point \(\lambda\) to the set
\(\Omega\), $0\notin \Omega$ with
\begin{equation}\label{reldist} \reldist(\lambda,\Omega) = 
\inf_{\mu  \in \Omega}\left|\frac{\mu - \lambda}{\mu}\right|.
\end{equation} 
Let \(\lambda \in \rho(H)\). Then $0\in \rho(G - \lambda J)$
and we will have \(\lambda \in \rho(H')\) if
\(Z^{-1}\) is bounded.
By Lemma \ref{lem:similarity} we have 
$1/\lambda \in \rho(G^{-1/2}JG^{-1/2})$ and hence we can write
\[
Z = (I - \lambda G^{-1/2}JG^{-1/2})(I + (I - \lambda G^{-1/2}JG^{-1/2})^{-1}\Delta).
\]
We claim that if $\lambda$ is such that $\kappa < \reldist (\lambda, \sigma(H))$ then $\lambda\in \rho(H')$.
This follows from the fact that  
\[
\|(I - \lambda G^{-1/2}JG^{-1/2})^{-1}\| = \frac{1}
{\reldist(\lambda,\sigma(H))}.
\]
Now, let $(\lambda^-,\lambda^+)$ be a spectral gap for $H$, i.e.\ $(\lambda^-,\lambda^+)\subset \rho(H)$. Then we have
\[
\reldist (\lambda, \sigma(H)) \le \min \left\{ \frac{|\lambda^- - \lambda|}{|\lambda^-|}, \frac{|\lambda^+ - \lambda|}{|\lambda^+|} \right\}
\]
and using this relation it is easy to prove the statements (i), (ii) and (iii) of the theorem.

To prove the analogous statement for the essential spectrum, first note that for $\lambda\in \rho(H')$ the resolvent of $H'$ can be written as
\begin{align}
  \label{eq:resolv}
  (H'-\lambda)^{-1} &= G^{-1/2}Z^{-1}G^{-1/2}J \\
  &= G^{-1/2}(I + (I - \lambda G^{-1/2}JG^{-1/2})^{-1}\Delta)^{-1}(I - \lambda G^{-1/2}JG^{-1/2})^{-1}G^{-1/2}J.
\end{align}
By 
$\mathcal{C}(\mathcal{X})$ define the Calkin algebra
$\mathcal{B}(\mathcal{X})/\mathcal{K}(\mathcal{X})$ where $\mathcal{B}(\mathcal{X})$ is the space of bounded operators on $\mathcal{X}$ and $\mathcal{K}(\mathcal{X})$ 
is the ideal of all compact operators. By 
\[
\hat{} \quad : \ \mathcal{B}(\mathcal{X}) \to \mathcal{C}(\mathcal{X})
\]
denote the corresponding Calkin homomorphism. 
We define a complex function $r$ with values in $\mathcal{C}(\mathcal{X})$ by
\begin{align*} 
r(\lambda) & = \reallywidehat{(H'-\lambda)^{-1}} \\ 
&= \reallywidehat{G^{-1/2}}(1 + (1 - \lambda \reallywidehat{G^{-1/2}JG^{-1/2}})^{-1}\hat\Delta)^{-1}(1 - \lambda \reallywidehat{G^{-1/2}JG^{-1/2}})^{-1}\reallywidehat{G^{-1/2}}\hat J.
\end{align*}
Since $r$ fulfils the resolvent equation it is a pseudoresolvent. From \cite[Theorem 5.8.6]{HP} it follows that $r$ admits a unique extension $\tilde r$ to a maximal pseudoresolvent. Since $0$ is an element of the domain of $r$, by \cite[Theorem 5.8.4]{HP} the domain of $\tilde r$ is the set of all $\lambda \in \mathbb{C}$ for which $1 - \lambda r(0)$ is regular, which is $\rho(r(0))^{-1}$. Hence the domain of $\tilde r$ is given by $\mathbb{C} \setminus \sigma(r(0))^{-1} = \mathbb{C} \setminus \sigma_{\mathrm{ess}}(H'^{-1})^{-1} = \mathbb{C} \setminus \sigma_{\mathrm{ess}}(H')$. 

Note that for selfadjoint elements in $\mathcal{C}(\mathcal{X})$ we have functional calculus and the norm of a selfadjoint element is equal to its spectral radius. Since $\reallywidehat{G^{-1/2}JG^{-1/2}}$ is a selfadjoint element in $\mathcal{C}(\mathcal{X})$, using these facts and Lemma \ref{lem:similarity} we obtain 
\begin{align*}
  \lVert (1 - \lambda \reallywidehat{G^{-1/2}JG^{-1/2}})^{-1} \rVert &= \frac{1}{\lvert \lambda \rvert } \left\lVert \left(\frac{1}{\lambda} - \reallywidehat{G^{-1/2}JG^{-1/2}}\right)^{-1} \right\rVert \\
  &= \left(\lvert \lambda \rvert  \mathop{\mathrm{dist}}\left( \frac{1}{\lambda},\sigma \left( \reallywidehat{G^{-1/2}JG^{-1/2}} \right)  \right)  \right)^{-1} \\
  &= \left(\lvert \lambda \rvert  \mathop{\mathrm{dist}}\left( \frac{1}{\lambda},\sigma_{\mathrm{ess}} \left( G^{-1/2}JG^{-1/2}\ \right)  \right)  \right)^{-1} \\
  &= \reldist (\lambda, \sigma_{\mathrm{ess}}(H))^{-1}.
\end{align*}
If $\lambda \in \mathbb{C} \setminus \sigma_{\mathrm{ess}}(H)$ and $\kappa < \reldist(\lambda, \sigma_{\mathrm{ess}}(H))$, from $\lVert \hat\Delta  \rVert  \le \lVert  \Delta \rVert < \reldist (\lambda, \sigma_{\mathrm{ess}}(H))$ it follows that $\lambda$ is in the domain of $r$ and hence also of $\tilde r$ which implies $\lambda \in \mathbb{C} \setminus \sigma_{\mathrm{ess}}(H')$. Now, to prove (iv), (v) and (vi), we can follow the same steps as in the proof of (i), (ii) and (iii). 
 \end{proof}
%



%
\begin{corollary}\label{GGdominate}
	Let $H$ be $j$-quasi-Hermitian with its polar decomposition $H = JG$ and let $ \tilde G \ge G$ and $\tilde H = J\tilde G$. 
	Then we have:
	\begin{enumerate}
		\item if $(\lambda_-,\lambda_+)\subset \rho(H)$ and 
	$0\in (\lambda_-,\lambda_+)$, then $(\lambda_-,\lambda_+)\subset \rho(\tilde H)$,
		\item if $(\sigma_-,\sigma_+)\cap \sigma_{\mathrm{ess}}(H) = \emptyset$ and 
	$0\in (\sigma_-,\sigma_+)$, then $(\sigma_-,\sigma_+)\cap \sigma_{\mathrm{ess}}(\tilde H) = \emptyset$.
	\end{enumerate}
	Hence, the spectral gap (both standard and essential) of $H$ around zero is contained in the one of $\tilde H$.
\end{corollary}
\begin{proof}
	Let us prove 1. First note that $G > 0$ and $(\lambda_-,\lambda_+)\subset \rho(H)$  implies $G - \lambda J > 0$ for all $\lambda \in (\lambda_-,\lambda_+)$.  Let $\lambda \in (\lambda_-,\lambda_+)$ be arbitrary. Then $0\in \rho(H - \lambda I)$ which implies $0 \in \rho (G - \lambda J)$. From the assumption it follows $0 < G - \lambda J \le \tilde G - \lambda J$, hence we have $0 \in \rho (\tilde G - \lambda J)$, which implies $\lambda \in \rho (\tilde H)$.

	To prove 2.\ note that $\sigma(H)$ contains in $(\sigma_-,\sigma_+)$ at most countably many eigenvalues taking into account its multiplicities. Let $\psi_k^{\pm}$, $k\in I_{\pm}$ be eigenvectors corresponding to positive/negative eigenvalues $\lambda_k^\pm$ of $H$ in $(\sigma_-,\sigma_+)$ chosen to be $J$-orthonormal. We define  
	\[ P=  \sum_{k\in I_+} (\sigma_+ - \lambda_k^+)(\cdot, J \psi_k^+) \psi_k^+ + \sum_{k\in I_-} (\sigma_- - \lambda_k^-)(\cdot, J \psi_k^-) \psi_k^- . \] 
	By the property $\lambda_k^\pm \to \sigma_\pm$ in the case when there are infinitely many positive/negative eigenvalues, the operator $P$ is $J$-selfadjoint, compact, $JP \ge 0$ and $(\sigma_-,\sigma_+) 
	\cap \sigma(H + P) = \emptyset$. Analogously as above, the last relation implies $0\in \rho (G + JP -\lambda J)$ for all $\lambda \in (\sigma_-,\sigma_+)$ and by the same reasoning as above we have  $0 <  G + JP - \lambda J \le \tilde G + JP -\lambda J$. Hence $\lambda \in \rho(\tilde H + P)$ for all $\lambda \in (\sigma_-,\sigma_+)$ and the claim now follows from the fact that the essential spectrum is invariant with respect to a compact perturbation.  
\end{proof}
In fact, the whole spectrum is expected to move asunder as \(G\) grows.
More specifically, by setting
\begin{equation}\label{deltag_pm}
\kappa_- = \inf_\psi \frac{\delta g(\psi,\psi)}{g(\psi,\psi)},
\quad 
\kappa_+ = \sup_\psi \frac{\delta g(\psi,\psi)}{g(\psi,\psi)},
\quad \kappa =\max\{|\kappa_-|,|\kappa_+|\},
\end{equation}
$\mathcal{I} = ((1+\kappa)\lambda_-, (1 + \kappa)\lambda_+)$ is expected 
to be replaced by
\begin{equation}\label{I+pm}
\mathcal{I}_+ = ((1 + \kappa_-)\lambda_-, (1 + \kappa_-)\lambda_+)
\end{equation}
(and analogously for the essential spectrum).
A general proof of this is still not available but there is a simple
rescaling construction which
improves the existing estimates in this direction
and has an independent interest besides.
We start from \eqref{deltag_pm} such that \(\kappa_\pm\) are finite
and
\begin{equation}\label{kappa-1}
\kappa_- > -1
\end{equation}
(this is anyhow necessary to guarantee the positive definiteness
of \(g + \delta g\)).
Now, for any \(\kappa_0 > -1\) we may write
\begin{equation}\label{rescaling}
g + \delta g = (1 + \kappa_0)g + \delta g',\quad
\delta g' = \delta g - \kappa_0g.
\end{equation}
The form \((1 + \kappa_0)g \) (being still positive definite)
yields the operator \((1 + \kappa_0)H \)
with the spectrum accordingly stretched and
\begin{equation}\label{deltagk'}
|\delta g'| \leq \kappa'(1 + \kappa_0)g, \quad
\kappa' = \max\left\{\frac{|\kappa_- - \kappa_0|}{1 + \kappa_0},
\frac{|\kappa_+ - \kappa_0|}{1 + \kappa_0}\right\}.
\end{equation}
Now we can optimise \(\kappa'\) in dependence of \(\kappa_0\).
The best value is obviously
\begin{equation}\label{kappa0opt}
\kappa_0 = \hat{\kappa}_0 = \frac{\kappa_+ + \kappa_-}{2},
\end{equation}
thus yielding
\begin{equation}\label{kappa'opt}
\hat{\kappa}' = \frac{\kappa_+ - \kappa_-}{2 + \kappa_+ + \kappa_-}.
\end{equation}
A first observation is that under the mere condition \(\kappa_- > -1\)
we already have \(\hat{\kappa}' < 1\) !
This means that the condition 
\eqref{kappa-1} automatically insures the statements of both Theorem
\ref{spectrality} (for the form \(v\))
and Theorems \ref{stability} and \ref{theorem:inclu} 
(for the form \(\delta g\)).\footnote{While the parameter
\(\mu\) in Theorem
\ref{spectrality} 
and Corollary \ref{cor:gap} may be 
called the (additive) spectral shift, the
parameter \(\kappa_0\) can analogously be called the
'multiplicative' spectral shift.} And all this 
{\em without any further assumption on the size of \(\kappa_+\)}.

Furthermore a simple inspection shows that we always have
\(\hat{\kappa}' \leq \kappa\), the equality being taken if and only if
\(\kappa_- = - \kappa_+\). 
Thus, (ii) in Theorem \ref{theorem:inclu} can be replaced by
\begin{equation}\label{Ipm1}
(1 + \hat{\kappa}_0) ((1 + \hat{\kappa}')\lambda_-, (1 - \hat{\kappa}'
)\lambda_+) \subseteq \rho(H')
\end{equation}

The improvement is particularly drastic if
\(\kappa_-, \kappa_+\) are close. The same improvement holds
for the essential spectral bound in Theorem \ref{theorem:inclu}.

The construction \eqref{rescaling} makes it possible to give
a natural strengthening of Corollary \ref{GGdominate}.
\begin{corollary}\label{GGGdominate}
If in Corollary \ref{GGdominate} the operators \(G\) and \(\tilde G\)
have the same form domains then \(\tilde H\) is $j$-quasi-Hermitian if and only if \( H\) is $j$-quasi-Hermitian.
\end{corollary}
\begin{proof} The same form domain means the double inequality
\(g \leq \tilde g \leq (\kappa_+ + 1)g\) for some \(\kappa_+ \geq 0\).
Then for \(\delta g = \tilde g - g\) we have 
\(0 \leq \delta g \leq (\kappa_+ - 1)g\) and our rescaling
construction \eqref{rescaling} gives
\[
|\delta  g| \leq \frac{\kappa_+ - 1}{\kappa_+ + 1}
\left(1 + \frac{\kappa_+ - 1}{2}\right)g 
\]
and Theorem \ref{stability} immediately applies.
 \end{proof}

\subsubsection*{Norm bounds} 
\label{ssub:norm_bounds}

If the form \(v\) in Theorem \ref{spectrality}
is just bounded:
\[
|v(\psi,\psi)| \leq a(\psi,\psi)
\]
for some positive \(a\) then in
\eqref{deltag} we obtain
\(\kappa \leq a\|G^{-1}\|\) and the typical inclusion from
Theorem \ref{theorem:inclu}, say, (i), becomes
\begin{equation}\label{(i)'}
((1+a\|G^{-1}\|)\lambda^-,(1 - a\|G^{-1}\|)\lambda^+)\subset \rho(H')
\end{equation}
which is quite pessimistic, if \(\lambda_+\) is large that is,
if the spectral gap is far away from zero. In fact, unnecessarily 
pessimistic, since in this case we have \(H'= H + S\), where \(S\) is the bounded operator generated by 
the form \(v\) with \(\|S\| \leq a\).
An alternative is to use the fact that 
the operator $H$ is selfadjoint with the respect to the scalar product 
\[
\langle \psi, \phi \rangle = (J J_1 \psi, \phi), 
\]
where $J_1 = \sign H$. Hence the operator
\[
\hat{H} = (JJ_1)^{1/2}H(JJ_1)^{-1/2}
\]
is selfadjoint in \(\mathcal{X}\). Then 
\[
\hat{H}' = (JJ_1)^{1/2}H'(JJ_1)^{-1/2} = \hat{H} + \hat{S}
\]
where 
\[
\hat{S} = (JJ_1)^{1/2}S(JJ_1)^{-1/2} \mbox{ and } 
\|\hat{S}\| \leq \|J_1\|a,
\]
so we have to do with a bounded perturbation of a selfadjoint operator
and 
 \eqref{(i)'} is replaced
by the uniform estimate
\begin{equation}\label{new(i)}
(\lambda^- + a\|J_1\|,\lambda^+ - a\|J_1\|)\subset \rho(H').
\end{equation}
which is independent of the size of \(\lambda_+\).

In the case of a Klein-Gordon Hamiltonian
\(\mathbf{H} = \mathbf{J}\mathbf{G}\) with \(\mathbf{G}\) from 
\eqref{G_factor} we cannot expect that the mere boundedness
of the operator \(V\) insures the same for the form
\(\mathfrak{v}\) in \eqref{vv}. The most interesting situation is the one
with \(\mathcal{X} = L^2(\mathbb{R}^n)\), $U^2 = - \Delta + m^2$ and $V$ 
is the multiplication operator. Then the condition that \(U^{1/2}VU^{-1/2}\)
is 
bounded is equivalent to the claim that $V$ is bounded multiplier 
in the space $W_2^{1/2}(\mathbb{R}^n)$. The characterization of 
such multipliers is given in \cite{MazyaShap}. This is not easy to check 
except for concrete functions $V$ (cf.~\cite{v0old}).   

Now we go over to two-sided estimates. They will be derived from corresponding
minimax formulae (cf.\ e.g.\ \cite{LangerNajmanTretter2006}), which we will 're-derive' in the process of proving the following result.
\begin{theorem}
 	\label{theorem:varchar}
Let $G$, $H$, $J$, $g$, $\delta g$, $ \kappa$, $H'$ be as in Theorem 
\ref{stability}. Let us denote by $\sigma^\pm$ ($\sigma'^{\pm}$) 
the infimum/supremum of positive/negative essential spectrum of the 
operator $H$ ($H'$), and by 
$\lambda_i^\pm$ ($\lambda_i'^\pm$) $i\in \mathbb{N}$ the 
positive/negative eigenvalues of $H$ ($H'$) ordered in the 
increasing/decreasing way, if positive/negative eigenvalues are 
smaller/larger than $\sigma^\pm$ ($\sigma'^\pm$) and 
$\lambda_i^\pm = \sigma^\pm$  ($\lambda_i'^\pm = \sigma'^\pm$) 
otherwise.  Then we have
\begin{equation}\label{eq:lam_lam'}
\kappa_- \le \frac{\lambda_k'^+ - \lambda_k^+}{\lambda_k^+} 
\le \kappa_+ \text{ for all } k\in \mathbb{N},
\end{equation}
with $\kappa_\pm$ as in \eqref{deltag_pm}, and similarly for $\lambda_k'^-$, $\lambda_k^-$. 

Especially, if $n \in \mathbb{N}$ is such that 
$(1 -\kappa)\lambda_n^+ \le (1 - \kappa)\sigma^+$ then
$\lambda_k'^+$, $k=1,\ldots,n$ are isolated eigenvalues of 
the operator $H'$ (always counted multiplicities), and analogously 
for negative eigenvalues.
\end{theorem}
\begin{proof}
From Lemma \ref{lem:similarity} we know that the operators ${H}^{-1}$ and ${H}'^{-1}$ are similar to the operators 
${G}^{-1/2}{J}{G}^{-1/2}$ and ${G}'^{-1/2}{J}{G}'^{-1/2}$, respectively. 
\begin{center}
\begin{tikzpicture}
\draw[latex-latex, thin, gray] (-3.5,0) -- (3.5,0) ; 
\draw[shift={(-2,0)},color=black] (0pt,0pt) -- (0pt,-3pt) node[below] {$\sigma^-$};
\draw[shift={(-1,0)},color=black]  node[below] {$\lambda_i^-$};
\draw[shift={(0,0)},color=black] (0pt,0pt) -- (0pt,-3pt) node[below] {$0$};
\draw[shift={(1,0)},color=black]  node[below] {$\lambda_i^+$};
\draw[shift={(2,0)},color=black] (0pt,0pt) -- (0pt,-3pt) node[below] {$\sigma^+$};
\draw[green,very thick] (-2,0) -- (-0.52,0);
\draw[orange,very thick] (-3.5,0) -- (-2,0);
\draw[green,very thick] (0.52,0) -- (2,0);
\draw[orange,very thick] (2,0) -- (3.5,0);
\end{tikzpicture}
\end{center}
Now we use the minimax theorem (c.f.\ \cite{Schm}) on the operator $-{G}^{-1/2}{J}{G}^{-1/2}$ to obtain for all $k\in \mathbb{N}$
\[
 -\frac{1}{\lambda_k^+} = \inf_{\substack{\mathcal{S}_k \\ \dimension \mathcal{S}_k = k}} \sup_{0\ne \psi\in \mathcal{S}_k} \frac{-({G}^{-1/2}{J}{G}^{-1/2}\psi,\psi)}{(\psi, \psi)}.
 \] 
This implies
\[
\frac{1}{\lambda_k^+} = \sup_{\substack{\mathcal{T}_k\subset D({G}^{1/2}) \\ \dimension \mathcal{T}_k = k }} \inf_{0\ne \varphi\in \mathcal{T}_k} \frac{({J}\varphi,\varphi)}{\lVert {G}^{1/2}\varphi \rVert^2 }.
\]
Since we know that $\lambda_k^+>0$ we have 
\[
\frac{1}{\lambda_k^+} = \sup_{\substack{\mathcal{T}_k\subset D({G}^{1/2})^+  \\ \dimension \mathcal{T}_k = k}} \inf_{0\ne \varphi\in \mathcal{T}_k} \frac{({J}\varphi,\varphi)}{\lVert {G}^{1/2}\varphi \rVert^2 },
\]
where $D({G}^{1/2})^+ = \left\{ \psi \in D({G}^{1/2})\colon ({J}\psi, \psi)>0 \right\}$. Now we can reverse the last equality and obtain 
\[
\lambda_k^+ = \inf_{\substack{\mathcal{T}_k\subset D({G}^{1/2})^+ \\ \dimension \mathcal{T}_k = k}} \sup_{0\ne \varphi\in \mathcal{T}_k} \frac{\lVert {G}^{1/2}\varphi \rVert^2}{({J}\varphi,\varphi)}.
\]
The same variational characterization holds for the eigenvalues $\lambda_k'^+$ of the operator ${H}'$
\[
\lambda_k'^+ = \inf_{\substack{\mathcal{T}_k\subset D({G}^{1/2})^+ \\ \dimension \mathcal{T}_k = k}} \sup_{0\ne \varphi\in \mathcal{T}_k} \frac{\lVert {G}'^{1/2}\varphi \rVert^2}{({J}\varphi,\varphi)},
\]
where we used the fact that $D({G}^{1/2})=D({G}'^{1/2})$.
Analogous formula holds for the negative eigenvalues of the operators ${H}$ and ${H}'$, just replace $H$ by $-H$ and $J$ by $-J$. 

Thus, $\lambda_k^+ $ depend monotonically on $G$ and we are in a position to use the estimate
\[
(1 + \kappa_-) g(\psi,\psi) \le g'(\psi,\psi) \le (1 + \kappa_+ ) g(\psi,\psi),
\]
hence \eqref{eq:lam_lam'} follows.

The other statement follows from the bounds 
$(1 - \kappa)\sigma^+ \le \sigma'^+$ and 
$\lambda_k'^+ \ge (1 + \kappa)\lambda_k^+$ and analogous ones for negative eigenvalues.
 \end{proof}
If only $\kappa$ from \eqref{deltag} is known then \eqref{eq:lam_lam'} gives
\begin{equation}
	\label{eq:lam_lam_kappa}
	\frac{|\lambda_k'^\pm - \lambda_k^\pm|}{|\lambda_k^\pm|} \le \kappa.
\end{equation}

A typical situation is to have \(H_0,H,G,G_0,v\) as in Theorem \ref{spectrality}
and then add a perturbation \(\delta g\), measured by \(G_0\). We have the following result.
\begin{corollary}\label{vdeltag}
Let \(H,G,G_0,v,b\) be as in Theorem \ref{spectrality}, assume that $H_0$ is selfadjoint, and let
\(\delta g\) be a symmetric form satisfying
\begin{equation}\label{c}
|\delta g(\psi,\psi)\| \leq c\|G_0^{1/2}\psi\|^2,\quad c + b < 1
\end{equation}
Then \eqref{deltag} holds with
\begin{equation}
\label{kappacb}
\kappa = \frac{c}{1-b} < 1.
\end{equation}
\end{corollary}
\begin{proof} Using \eqref{GleqG0} we have
\[
|\delta g(\psi,\psi)\| \leq c\|G_0^{1/2}\psi\|^2 
\leq \frac{c}{1-b}g(\psi,\psi). \qedhere 
\]
 \end{proof}

All quantitative bounds obtained thus far can be immediately
extended to the operators satisfying the condition A.3 with a general
\(\lambda\) thus giving new  bounds which could then be optimised
over \(\lambda\), if so desired.

\section{Perturbations of spectra, Klein-Gordon structure}
Theorem \ref{spectrality} and Corollary \ref{cor:gap}
and all subsequent results, in particular Theorems \ref{theorem:inclu}
and \ref{theorem:varchar}, are immediately applied to the
Klein-Gordon operator with the potential \(V\) perturbed into
\(V + \delta V\) where \(\delta V\) is again symmetric
and defined  on
\(\mathcal{D}(U)\) with
\begin{equation}\label{cb1}
c = \|\delta VU^{-1}\|,\quad c + b < 1.
\end{equation}
So, for instance, \eqref{eq:lam_lam_kappa} becomes 
\begin{equation}\label{lam_lam'1}
 \frac{|\lambda_k^{\prime\pm} - \lambda_k^{\pm}|}
{|\lambda_k^\pm|} \le \frac{c}{1 - b}
\end{equation}
and so on.

The matrix structure of the Klein-Gordon operator allows for
somewhat improved bounds.

Set $A = V U^{-1}$, $\delta A = \delta V U^{-1}$, where
\(\delta V\) is symmetric and \(V\) is perturbed into
\(V + \delta V\). Assume
\begin{equation}\label{deltaAbc} 
\|A\| \leq b, \|\delta A\| \leq c,\quad
c + b < 1,
\end{equation}
By \eqref{vv} and \eqref{vv<bU} the form 
\(\mathfrak{g} = \mathfrak{g}_0 + \mathfrak{v}\)
can be written as
\[
\mathfrak{g}(\psi,\varphi) = 
(\mathbf{A}\mathbf{U}_0^{1/2}\psi,\mathbf{U}_0^{1/2}\varphi), \quad \psi, \varphi\in \mathcal{D}(\mathbf{U}_0^{1/2}).
\]
Set
\[
\delta \mathbf{A} =   
\begin{bmatrix}
        0 & \delta A^{\ast} \\ \delta A & 0
\end{bmatrix}.
\] 
Obviously \(\|\delta \mathbf{A}\| = \|\delta A\| \leq c\).
The form $\mathfrak{g}$ is perturbed into 
$\mathfrak{g} + \delta \mathfrak{g}$ with
\[
 \delta \mathfrak{g} (\psi, \varphi) = (\delta \mathbf{A} \mathbf{U}^{1/2}\psi,\mathbf{U}^{1/2}\varphi) \quad \psi, \varphi\in \mathcal{D}(\mathbf{U}_0^{1/2}). 
 \] 
Thus, we are again in the conditions of Theorem
\ref{spectrality}, applied to the form sum
\[
\mathfrak{g}' = \mathfrak{g}_0 + \mathfrak{v} + \delta\mathfrak{g}
\]
(recall that \(b + c  < 1\)). Since $\mathbf{U}$ is selfadjoint, the operator \(\mathbf{H}' = \mathbf{J} \mathbf{G}'\) is 
$j$-quasi-Hermitian. To estimate the perturbation we have to compute
\[
\frac{\delta\mathfrak{g}(\psi,\psi)}{\mathfrak{g}(\psi,\psi)} =
\frac{(\delta \mathbf{A}\mathbf{U}_0^{1/2}\psi,\mathbf{U}_0^{1/2}\psi)}
{(\mathbf{A}\mathbf{U}_0^{1/2}\psi,\mathbf{U}_0^{1/2}\psi)} =
\frac{(\delta \mathbf{A}\phi,\phi)}{(\mathbf{A}\phi,\phi)},
\]
where $\psi \in \mathcal{D}(\mathbf{U}_0^{1/2})$ and $\phi = \mathbf{U}_0^{1/2}\psi$.
To proceed further we use the 'block Cholesky' factorisation
\[
\mathbf{A} =
\begin{bmatrix}
(I - A^*A)^{1/2} & A^{\ast} \\  0 & I
\end{bmatrix}
\begin{bmatrix}
(I - A^*A)^{1/2} & 0 \\  A & I
\end{bmatrix}.
\]
Then
\[
\frac{\delta\mathfrak{g}(\psi,\psi)}{\mathfrak{g}(\psi,\psi)} =
\frac{(\mathbf{L}^*\delta\mathbf{A}\mathbf{L}\chi,\chi)}{(\chi,\chi)},
\]
where 
\[
\mathbf{L} = 
\begin{bmatrix}
(I - A^*A)^{-1/2} & 0 \\  -A(I - A^*A)^{-1/2} & I
\end{bmatrix}
\] 
is the inverse right Cholesky factor of \(\mathbf{A}\) above and $\chi = \mathbf{L}^{-1} \phi$.
The operator \(\mathbf{L}^{\ast} \delta \mathbf{A} \mathbf{L}\)
is bounded and it is given by the matrix
\[
\mathbf{M} =  \mathbf{L}^{\ast} \delta \mathbf{A} \mathbf{L} = 
\begin{bmatrix}
-(I- A^{\ast}A)^{-1/2}(\delta A^{\ast} A + A^{\ast} \delta A ) (I- A^{\ast}A)^{-1/2} & (I- A^{\ast}A)^{-1/2} \delta A^{\ast} \\
\delta A (I- A^{\ast}A)^{-1/2} & 0
\end{bmatrix},
\]
such that $\kappa_+ = \max \sigma(\mathbf{M})$, $\kappa_- = \min \sigma(\mathbf{M})$.
Let $ a_- I < -(I- A^{\ast}A)^{-1/2}(\delta A^{\ast} A + A^{\ast} \delta A ) (I- A^{\ast}A)^{-1/2} < a_+ I$ in the form sense and let 
\[
\mathbf{T}_\kappa = \begin{bmatrix}
a I & B^{\ast} \\ B & 0
\end{bmatrix},
\]
where $B=\delta A (I- A^{\ast}A)^{-1/2}$. Then 
\[
\mathbf{T}_{a_-} \le \mathbf{M} \le \mathbf{T}_{a_+},
\]
and to obtain bounds for $\frac{\lvert \delta \mathfrak{g}(\psi,\psi)\rvert }{\mathfrak{g}(\psi,\psi)}$ it is sufficient to find upper bound 
for the operator $\mathbf{T}_{a_-}$ and a lower bound for $\mathbf{T}_{a_+}$.

A simplest way is to find just a norm bound.
Since $\mathbf{T}_{-a}$ is unitarily similar to $- \mathbf{T}_{a}$, it is enough to find a bound on $\max \sigma(\mathbf{T}_a)$. Let $\lambda \in \sigma(\mathbf{T}_a)$ be such that $\lambda = \lVert \mathbf{T}_a \rVert $. Then, since $\lambda$ is from the approximative point spectrum of the operator $\mathbf{T}_a$, there exists a Weyl sequence $\psi_n = \left( \begin{smallmatrix} \psi^1_n \\ \psi^2_n		
\end{smallmatrix} \right) $, $\lVert \psi \rVert = 1 $ such that $(\mathbf{T}_a - \lambda \mathbf{I})\psi_n \to 0$. It follows 
\[\lambda^2 \lVert \psi_n^1 \rVert^2 - \lambda a \lVert \psi_n^1 \rVert^2 - \lVert B \psi_n^1 \rVert^2 \to 0 \] 
and $\psi_n^1$ does not converge to zero. Hence 
\[
\frac{a \lVert \psi_n^1 \rVert + \sqrt{a^2 \lVert \psi_n^1 \rVert^2 + 4 \lVert B\psi_n^1 \rVert^2 } }{2\lVert \psi_n^1 \rVert} \to \max \sigma(\mathbf{T}_a) = \lVert \mathbf{T}_a \rVert. 
\]
This implies $\lVert \mathbf{T}_a \rVert \le \frac{a + \sqrt{a^2 + \lVert B \rVert^2 }}{2}$. Since $\lVert B \rVert \le \frac{\lVert \delta A \rVert }{\sqrt{1-b^2}}$, we obtain 
\[
\frac{\lvert \delta \mathfrak{g}(\psi,\psi)\rvert }{\mathfrak{g}(\psi,\psi)} \le \frac{1}{2} a + \sqrt{\left(\frac{a}{2}\right)^2 + \frac{\lVert \delta A \rVert^2}{1-b^2}}. 
\]
If we take $a= \frac{2b \lVert \delta A \rVert}{1-b^2}$, we 
retrieve the already obtained bound \eqref{kappacb}. If the
perturbation \(\delta V\) is measured by \(V\), that is,
\begin{equation}
\label{eq:deltaVV}
\|\delta V\psi\| \leq \nu\|V\psi\|,
\end{equation}
then by the analogous calculation \eqref{kappacb} is replaced by
\begin{equation}
\label{eq:kappanub}
\kappa = \frac{\nu b}{1 - b} < 1.
\end{equation}

If $V$ and $\delta V$ have ''disjoint supports'', 
meaning that 
\begin{equation} \label{disjoint}
 (\delta V \psi, V \psi) = 0  
\mbox{ for all} \psi \in D(U),
\end{equation}
 then the 1,1-block in
\(\mathbf{M}\) vanishes and
we obtain an improved bound
\begin{equation}
\label{eq:pert_estimate2}
\frac{\lvert \delta \mathfrak{g}(\psi,\psi)\rvert }
{\mathfrak{g}(\psi,\psi)} \le 
\kappa = \frac{\lVert \delta A \rVert}{\sqrt{1-b^2}}
= \frac{c}{\sqrt{1-b^2}} 
\end{equation}
where the requirement \(\kappa < 1\) is equivalent to
\[
b^2 + c^2 < 1.
\]
A weakening of the assumption \eqref{disjoint}
is to ask that $V \delta V$ be of same sign, say 
$V \delta V \leq 0$, meaning 
$\delta A^{\ast} A + A^{\ast} \delta A \le 0$. Now we can 
take $a_-=0$ and $a_+ = \frac{2b \lVert \delta A \rVert}{1-b^2}$. 
This implies
\begin{equation}
\label{eq:pert_estimate3}
- \frac{\lVert \delta A \rVert }{\sqrt{1-b^2}} \le  - \lVert B \rVert = - \lVert \mathbf{T}_0 \rVert \le \frac{\delta \mathfrak{g}(\psi,\psi) }{\mathfrak{g}(\psi,\psi)} \le \lVert \mathbf{T}_{a_+} \rVert \le \frac{\lVert \delta A \rVert }{1-b}.
\end{equation}
This leads to \(\kappa_\pm\) in \eqref{eq:lam_lam'} with
\(\kappa_+ + \kappa_- \neq 0\) and an improved bound 
is obtained (we omit the details).
\section{Applications}

\begin{example}
This is an example which illustrates the general theory, 
developed in Section \ref{sec:perturbations_of_spectra}. We  set \(H = JG\) where \(G\) is the selfadjoint
realisation of \(- \mathrm{d}^2/\mathrm{d}x^2\) in \(L_2[-1,1]\) with Dirichlet 
boundary conditions and
\[
J\psi(x) = j(x)\psi(x)
\]
where \(j\) is a piecewise constant function with values in
\(\{-1,1\}\). The eigenvalue problem for H may be written as
\[
-\psi''(x) = \lambda j(x)\psi (x)
\]
so this is a simple Sturm-Liouville problem with 'indefinite weight'.
\'Curgus and Najman \cite{CurgusN1995operator} have proved that \(H\) is \(j\)-quasi-Hermitian.
The spectrum consists of simple eigenvalues which accumulate in \(\pm\)
infinity. Now set
\[
H' = JG',\quad G' = -\frac{\mathrm{d}}{\mathrm{d}x}\left((1 + \delta)\frac{\mathrm{d}}{\mathrm{d}x}\right)
\] 
with \( \delta\) measurable and \(\sup_x|\delta(x)| < 1\).
Here we have
\[
\delta g(\psi,\phi) = \int\delta(x)\psi'(x)\overline{\phi'(x)}\,\mathrm{d}x
\]
and consequently
\[
\kappa_+ = \sup_x\delta(x),\quad \kappa_- = \inf_x\delta(x).
\]
Theorem \ref{theorem:varchar} immediately gives
\[
\inf_x\delta(x) \leq \frac{\lambda_k^{\prime +} - \lambda_k^{+}}{\lambda_k^{+}}
\leq \sup_x\delta(x)
\]
and similarly for \(\lambda_k^{\prime -}\). For \(j \equiv 1\) this gives
the same bound as in the standard selfadjoint case. Note that our bounds 
are always independent of the spectral condition of the operator \(H\)! This phenomenon was observed in \cite{vs} for finite matrices.
Our results are immediately applicable to the Laplacian in higher
dimansions.
\end{example}
\begin{remark}
One can also apply the theory developed in Section \ref{sec:perturbations_of_spectra} to various classes of differential eigenvalue problems where the corresponding operator  can be written in the form $H=JG$ with $J$ a reflection and $H$ similar to a selfadjoint operator. Such operators arise, for example, in the case of
\begin{itemize}
	\item the weigthed eigenvalue problem $Lu = \lambda (\sgn x_n) u$ with the domain $\mathbb{R}^n$ where $L$ is positive symmetric partial differential operator with constant coefficients, see \cite{CurgusN1998positive},
	\item the elliptic eigenvalue problem with indefinite weight $(- \Delta + 1 + q)u = \lambda r u$ in $\mathbb{R}^n$ where $r$ changes sign in $\mathbb{R}^n$ and satisfies some additional assumptions and $q$ is relatively compact with respect to $- \Delta$, see \cite{CurgusN1994},
	\item the operator $\frac{\sgn x}{\lvert x \rvert^\alpha }\left( - \frac{\mathrm{d}^2}{\mathrm{d} x^2} + c \delta \right) $ with $\alpha > -1$ and $\delta$ the Dirac delta, see \cite{Kostenko2006},
	\item the one-dimensional Schr\"{o}dinger-type operators in a bounded interval with non-self-adjoint Robin-type boundary conditions, see \cite{Krejcirkk2014}, and
	\item the operator arising in the study of stability of solitons for the 1-D relativistic Ginzburg–Landau equation, see \cite{Kostenko2015}.
\end{itemize}
However, perturbing the indefinite weight seems to be a more delicate
matter cf.\ \cite{NajmanV1998}.
\end{remark}
\begin{example}
Consider the Klein Gordon operator \(\mathbf{H}\) satisfying the conditions A.1-A.3. We consider here 
a bounded perturbation \(\delta V\).
Then in \eqref{cb1} we have 
\(c \leq \|\delta V\|\|U^{-1}\|\) 
and from \eqref{lam_lam'1} we have 
\begin{equation}
\label{eq:h_bound}
|\lambda_k^{\prime\pm} - \lambda_k^\pm| \leq h\|\delta V\|
\end{equation}
with 
\[
h = \frac{|\lambda_k^\pm|\|U^{-1}\|}{1 - b}.
\]
By taking for the moment \(\delta V = \delta I\) (a scalar multiple of the identity) then
we have exactly
\(\lambda_k^{\prime\pm} - \lambda_k^\pm = \delta\)
so that the factor \(h\)  is never less than \(1\). It will
attain \(1\), if \(b = 0\) and \(|\lambda_k^\pm|\|U^{-1}\| = 1 \), that is, 
\(V = 0\) and \(\lambda_k^\pm\) is just a ground state. As can be 
expected with
relative estimates, ours will be poor for absolutely large eigenvalues,
but these are seldom interesting in practice.

If there is more information on \(\delta V\) then the bounds can be tightened.
Set
\[
\delta_- = \inf \delta V,\quad \delta_+ = \sup \delta V
\]
and write
\[
\mathbf{H}' = \mathbf{J}\mathbf{G}' = \mathbf{J}(\mathbf{G} + \delta \mathfrak{g}) = \mu \mathbf{I} + \mathbf{J}(\mathbf{G} + \delta \mathfrak{g} - \mu \mathbf{J}).
\]
So, it is enough to consider the perturbation \(\mathbf{G} \to \mathbf{G} + \delta \mathfrak{g} - \mu \mathbf{J}\)
and then just shift for \(\mu\). Now \(\delta \mathfrak{g} - \mu \mathbf{J}\) is produced by the
potential \(\delta V -  \mu I\) and \eqref{eq:h_bound} gives here
\[
|\lambda_k^{\prime\pm} - \lambda_k^\pm| \leq h\|\delta V -  \mu I\|.
\]
Obviously a best \(\mu\) is 
\[
\mu = \frac{\delta_+ + \delta_-}{2}
\]
thus giving
\begin{equation}\label{h_boundmu}
|\lambda_k^{\prime\pm} - \lambda_k^\pm| \leq h\frac{\delta_+ - \delta_-}
{2},
\end{equation}
a property analogous to the known ones for the Schr\"odinger or the Dirac 
operator.
\end{example}
\begin{example}
As a next illustrating model we consider the Hamiltonian given
by 
\begin{equation}\label{Hoscill}
\mathcal{X} = \mathbf{L}^2(\mathbb{R}),\quad
U^2 = \overline{-d^2/dx^2 + x^2 + \beta},\quad
V = V(x) = \alpha x,\quad \alpha,\beta \geq 0, \alpha < 1,
\end{equation}
that is, we have mass-term harmonic oscillator combined 
with the homogeneous electric field. 
The eigenvalues of the corresponding Klein-Gordon operator for the $n$-dimensional case can be computed exactly. They are
\begin{equation}\label{mu_harm}
\mu = \mu_n^\pm(\alpha) = \pm \sqrt{(1 - \alpha^2)\beta + 
(1 - \alpha^2)^{3/2}\sqrt{1 + 2n}}.
\end{equation}
Here to each eigenvector there correspond two different eigenvalues
but the formulae \eqref{linearisation} give two different eigenvectors
of \(H\), moreover, as we know from Theorem \ref{spectrality}, all those
eigenvectors {\em form a Riesz basis} of the Hilbert space 
\(\mathcal{X}\oplus \mathcal{X}\). In any case our bound
\(\|VU^{-1}\| < \alpha\) is shown to be sharp because  
at 
\[
\alpha = 1
\]
all eigenvalues tend to zero and in the same time they fill up
the whole real axis whereas
the eigenfunctions 
become flat and shift to infinity.

One can show that the situation is the same also in 
higher dimensions, the only difference is that then in
\eqref{mu_harm} the eigenvalues 
\(\mu_k^\pm\) will be accordingly multiple.

Now for perturbations. To stay within exact solvability
for \(V = \alpha x\) we set 
\[
\delta V = \epsilon V = \epsilon \alpha x
\]
and choose \(\beta = 0\)
Then our relevant bound \eqref{deltaAbc} 
has \(c =\epsilon\) and \(b = \alpha\), so \eqref{kappacb} gives
\[
\frac{|\mu(\alpha +\epsilon\alpha) - \mu(\alpha)|}{\mu(\alpha)}
\leq \frac{\epsilon}{1 - \alpha}
\]
whereas the exact formula gives, for \(\epsilon\) small\footnote{We consider
positive eigenvalues and omit their subscripts and superscripts.},
\[
\frac{\mu(\alpha +\epsilon\alpha) - \mu(\alpha)}{\mu(\alpha)}
\approx
\frac{\mu'(\alpha)}{\mu(\alpha)}\epsilon 
=
-\frac{3}{2}\frac{\alpha\epsilon}{1 - \alpha^2}
\]
\[
=
-\frac{3}{2}\frac{\alpha}{1 + \alpha}\frac{\epsilon}{1 - \alpha}
\]
which is absolutely smaller by the factor
\[
\frac{3}{2}\frac{\alpha}{1 + \alpha}
\]
which varies from zero to \(3/4\). Our estimate is poor at small
alphas and no wonder because at \(\alpha = 0\) the change of the eigenvalues
is quadratically small and ours is, by its nature, a 'first order estimate'. To capture
such behaviours other methods will be needed. This will be considered in a forthcoming paper. 

Anyhow since the bound \(\|VU^{-1}\| < 1\) has shown to be sharp
there is no need of considering 'shifted bounds' as in A.3.

\end{example}

The following example can be seen as a two dimensional
discretisation of the well-known square well potential Hamiltonian considered
in \cite{SSW}, and it has very much the same behaviour.

\begin{example}
Set
\[
U^2 =
\begin{bmatrix}
2 & -1   \\
-1   &  2 \\
\end{bmatrix},
\quad
V = \tau
\begin{bmatrix}
-1 & 0   \\
 0 & 0 \\
\end{bmatrix}
\]
with \(0 \leq \tau \leq 2\).
The behaviour of the spectrum as a function of \(t\)
is shown on Figure \ref{fig1}.

\begin{figure}[!htb]
\centering
\includegraphics*[width=8cm]{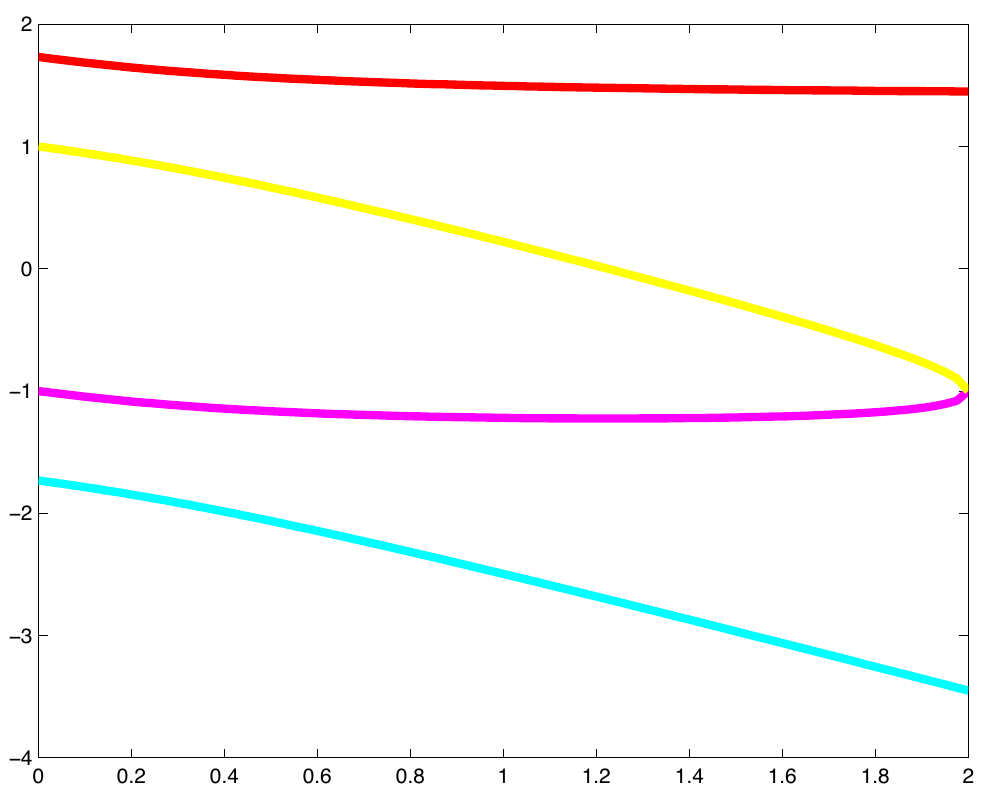}
\caption{Eigenvalues as functions of \(\tau \in [0,2]\)}
 \label{fig1}
\end{figure}

In fact, as the figure indicates, the operator \(H\) is well-behaved
for \( \tau < 2\) whereas at \( \tau = 2\) the two inner eigenvalues
\(\lambda_1^-\) and \(\lambda_1^+\) are equal
and for \(\tau > 2\) they become non-real. Indeed,
we have
\[
Q_\tau(\lambda) = 
\begin{bmatrix}
2 - (\lambda + t)^2 &   -1  \\
-1          & 2 - \lambda^2 \\
\end{bmatrix}
\] 
with 
\[
Q_2(-1) =
\begin{bmatrix}
1  &  -1  \\
-1 &   1 \\
\end{bmatrix},
\]
which has a one-dimensional null space. Thus, the eigenvalue \(-1\) is 
defective (non-semisimple).
This reproduces the phenomenon of
'Schiff-Snyder-Weinberg complex pair creation' observed in
\cite{SSW}.

The value $\lVert V U^{-1} \rVert $ here (numerically rounded) equals
\(0.745\, \tau\) which covers a bit more than
one half of the region of well-behavedness. Taking in 
\eqref{muVb} \(\mu = -\tau/2\) 
we obtain \(b = b_\tau = \tau/2 < 1\)
for \(\tau < 2\) which is sharp because at \(\tau = 0\) there is no
eigenbasis as was seen above.

We now illustrate some perturbations. 
For unperturbed operators we take
\(\tau = 0,1,1.8\) and for the perturbations
\[
\delta V = 
\begin{bmatrix}
\eta  &  0  \\
0     &   0 \\
\end{bmatrix}
\]
with \(\eta = 0.001, 0.1, 0.5\) thus obtaining altogether
nine examples. Tabulated below are the true maximal relative distances
\[
\max_i\frac{|\mu_i' - \mu_i|}{|\mu_i + \tau/2|}
\]
and our respective bounds
\begin{equation}\label{shifted_bound}
\kappa = \frac{\eta}{1 - \tau/2}
\end{equation}
which are derived from \eqref{eq:lam_lam_kappa} and  taken
for the shifted operator \(H + \tau/2 I\) according to what is
said in Remark \ref{shift}.

The true maximal relative distances are:

\begin{center}
\begin{tabular}{l|r|r|r}
        &   \(\eta\) = 0.001 & \(\eta\) = 0.1 & \(\eta\) = 0.3 \\ \hline
t = 0   &   5.0037e-04       & 5.3732e-02     & 1.8241e-01 \\
t = 1   &   1.3269e-03       & 1.3409e-01     & 4.1064e-01 \\
t = 1.7 &   3.3731e-03       & 3.4990e-01     & 1.4355e+00  \\
\end{tabular}
\end{center}

The respective bounds are:

\begin{center}
\begin{tabular}{l|r|r|r}
        &   \(\eta\) = 0.001 & \(\eta\) = 0.1 & \(\eta\) = 0.3 \\ \hline
t = 0   &   1e-03        &  1e-01      &  3e-01   \\
t = 1   &   2e-03        &  2e-01      &  6e-01   \\
t = 1.7 &   6.6667e-03        &  6.6667e-01      &  2e+00   \\
\end{tabular}
\end{center}
Similar results are obtained with randomly chosen perturbations \(\delta V\).
\end{example}

\noindent \textbf{Acknowledgements:}

\noindent The first author was partially supported by Croatian Science Foundation project grants 9345 and IP-2016-06-2468.

\bibliographystyle{plain}
\bibliography{kleingordon}

\end{document}